# Utilizing Symmetry of Planar Ultra-Wideband Antennas for Size Reduction and Enhanced Performance

Ahmed Toaha Mobashsher, Amin Abbosh

The School of Information Technology and Electrical Engineering

The University of Queensland

St. Lucia, Brisbane QLD 4072, Australia

E-mail: a.mobashsher@uq.edu.au, a.abbosh@uq.edu.au

*Abstract*— **With the increasingly new ultra-wideband applications, antenna researchers face huge challenges in designing novel operational geometries. Monopole and quasi-monopole antennas are seen to be the most compact and easily incorporable solution for portable devices taking the advantages of printed circuit board (PCB) techniques. Most antennas of such type have symmetrical structures. It is possible to attain wider operating bandwidths by meeting symmetry conditions while chopping the antenna into halves for a compact structure. However, there is no generalized way of applying such a technique. The presented paper addresses this issue by proposing a common feeding technique that can be applied to any antenna which is miniaturized using its symmetrical structure. The proposed technique enables feeding the halved structure to achieve wider and better impedance matching than the reported full-size antennas. The theory of characteristic modes is applied to quasi-monopole structures to get an insight of the antennas' mechanism. The radiation patterns are also correlated with modal current distributions to understand the radiation characteristics of the modified structure. Lastly, the method is implemented on some example antennas to illustrate its potential.**

## I. Introduction

ULTRA WIDEBAND (UWB) technology received tremendous attention from academia and industries. Although it was mainly used for military and radar applications since early 1960s, it has been widely utilized in recent years for the proliferation of short-range high-throughput wireless communications, medical imaging, ad-hoc networking, and other wireless sensing and monitoring applications [1]. The wide spread use of UWB technology is due to its prime advantages, such as high resolution reliable data, low transmission-power consumption, high immunity to multi-path interferences, high channel capacity and high security measures [2, 3].

By definition, a UWB signal should have more than 25% fractional bandwidth or occupy a bandwidth greater than 500 MHz [4]. The Federal Communication Commission allowed a frequency band from 3.1 GHz to 10.6 GHz for low power UWB communications. Other bands of operation are also defined in lower frequencies for some other pertinent UWB applications [5].

Antennas and wireless propagation community designed a huge number of UWB antennas even before the inception of UWB system [6, 7]. However, the increasingly new applications demanded novel features; antenna designers accepted the challenge and took the art of antenna design to a new level. Historically, three-dimensional antennas, such as bi-conical dipole or monopole [6, 8], spherical dipole [9] and various types of horn antennas [10, 11] were very popular to achieve wide bandwidth. The development history UWB antenna is adequately described in [7, 12, 13].

The breakthrough of printed circuit board (PCB) technique vigorously changed the evolution of UWB antenna design. Most of the three-dimensional antennas were translated into their two-dimensional versions (see [14] and [15]). The PCB technique created a new domain in the field of antenna research which resulted in low-cost and compact antenna structures that can be easily integrated into a circuit board and embedded into a portable device.

Both omni- and directional patterns can be achieved by the PCB fabricated antennas [16-19]. However, for the portable applications, omni-directionality is required to have surround reception and transmission of signals. Planar dipole antennas [20] and dipole-evolved monopole antennas [21-23] are mostly reported in the literature to attain improved operation of portable devices. Monopoles considerably decrease the required space to integrate into a communication device. UWB monopole antennas are primarily designed by meeting the electrical symmetry of dipole antennas. Afterwards, some miniaturization measures like imposing slots [24-26], truncations [27, 28], fractals [29-32], ground modifications [33-35], dielectric resonators [36], impedance transformers [37] or capacitively-loaded loop (CLL) resonators [38] are applied to realize smaller quasi-monopole UWB structures.

It is worth to mention that all the aforementioned miniaturization processes have some undesired effects on the radiation patterns generated by the antennas. Due to the planar orientation, the monopole antennas can be fed by microstrip lines, where the ground plane is situated at the opposite layer of the fed patch or by a co-planar-waveguide (CPW) structure, where all the antenna parts are printed on the same layer of the substrate. Fabricating the quasi-monopole antennas on high dielectric substrates can further miniaturize their size. However, the physical limitation still exists and the manufacturing cost also increases.

Most of the reported monopole and quasi-monopole UWB antennas have symmetrical structures [28-32, 34-41]. Numerous band-notch UWB quasi-monopole antennas have also illustrated symmetrical shapes [42-44]. In those structures, two strong current paths exist on two symmetrical halves. It is noted that the patterns of the current distribution matches the conditions of magnetic mirror symmetry. Some researchers have shown that chopping off half of the symmetrical monopole antenna retains one strong current path and can operate over a wider bandwidth [45-51]. Degradation in the cross-polarization level is reported as a limitation to this technique [52, 53]. However, not all half cut antennas can achieve wider bandwidth with suitable impedance matching right away. Hence, some literatures have reported that modifying feeding structure provides better impedance matching [48, 52, 54-57]. However, this method cannot be generalized since all these antennas are CPW fed, and microstrip fed beveled shaped UWB antenna failed to provide proper impedance matching with $50\Omega$ input system impedance [58]. The reason might be attributed to the high input impedance that halved microstrip feeding imposes on the antenna resulting in an unstable impedance matching.

Most of the reported literatures lack a productive physical insight of the beveled miniaturization technique. Lately, the use of the theory of characteristic modes (TCM) in explaining the physical meaning of antennas is becoming popular [59-61] and getting increasingly recognized by more researchers mostly for compact multiple-input multiple-output (MIMO) [62-65] and narrow/multi-band [66-69] antenna designs. TCM is simpler in explaining an arbitrary shaped structure compared to the full wave model or cavity model and provide adequate information to realize the mechanism behind antenna's performance [27]. Studying the antenna structure with the theory of characteristic modes is also becoming popular since it provides a better understanding of the wideband antenna's mechanism [70-75]. However, the reported TCM analysis of [76] does not correlate the altered radiation patterns with the modal currents when the UWB antenna is cut into halves.

The goal of this paper is to examine the effects of miniaturizing a planar antenna along the magnetic and/or electric boundary line starting from dipole to monopole antennas. A generalized way to improve the impedance matching of the monopole and quasi-monopole antennas with different feedings is also described with an explanation of the characteristic modes of antennas. Consequences of miniaturization on the current distributions, and thus the radiation patterns, are also described from the characteristic mode perspective. Performances of applying the new miniaturization technique are also compared with other variations. Finally, the new miniaturization technique is applied on some sample antennas with different feeding techniques to demonstrate the effectiveness of the proposed approach. Possible influential application of these miniaturized antennas is the integration into a compact UWB transceiver chip [77, 78].

## II. Boundary Conditions in a Typical Symmetrical Dipole Antenna

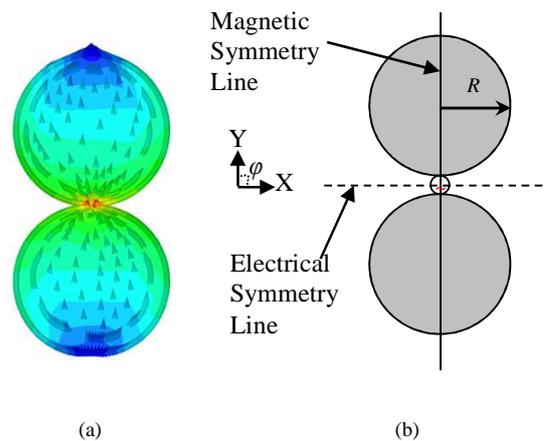

(a)                    (b)

Fig. 1. (a) Current distribution and (b) symmetry conditions of a typical dual disk dipole antenna.

A dual disk dipole (DDD) antenna, which represents a general shape of a dipole, in free space is shown in Fig. 1. The disks are of radius R = 15 *mm* and separated from each other by a distance 0.5 *mm*. It is to be noted that the selection of certain dimensions are needed in this and the following examples to show the performance of the example antennas using the full-wave electromagnetic simulators. That specific selection does not limit the application of the presented approach. The antenna is excited from the mid-point. Image theory of electromagnetics suggests that when an electric current source is placed next to an infinitely large electric conductor, the conductor can be substituted with an out of phase current source of the same magnitude [79, 80]. On the contrary, an electrical symmetry line exists between two identical electric fields with a 180° polarity difference relative to each other. Observing this criterion for the currents flowing on the dipole, we can identify the electrical symmetry condition between the two disks. The dashed line along X axis represents this symmetry of the dual disk dipole, keeping one disk in each side of the line. Similarly, the in-phase current distributions along Y axis evoke the in-phase magnetic fields on the dipole. Nevertheless, the magnitude is seen to be identical along the center line. Thus according to the image theory, a magnetic symmetry exists along the center line of Y axis [81]. The solid line represents the magnetic boundary condition which divides the dipole into two parts where each portion have two half circular disks.

Surface current distributions of an antenna at a certain frequency reveal the operating principle of the antenna at that particular frequency. As seen from Fig. 1, the currents are in mirror symmetry along the magnetic boundary line. Because of the symmetrical structure of the dual disk dipole, there exist two identical current paths in mirror. Each of those paths is reasonably enough to attain the resonance. This is why a dual half disk dipole (DHDD) antenna is constructed where the magnetic boundary condition is met to examine the effects on the antenna's performances after the cut. As anticipated, the direction of currents has not changed in the new DHDD antenna. However, when the DDD is shredded into two portions, the cut edge emits more radiation whereas the radiation from the middle part decreases. Thus, a significant amount of currents gather around the cut edge, while a slight decrement of current is observed in the inner portion.

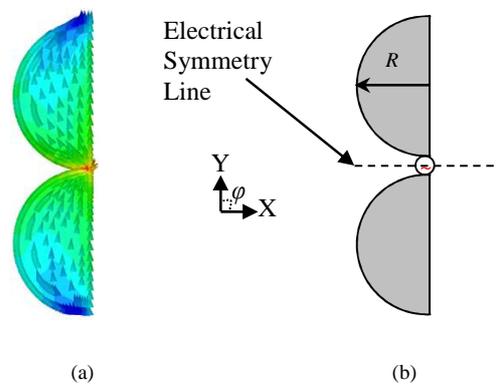

Fig. 2. (a) Current distribution and (b) symmetry condition of dual half disk dipole.

Cutting along the magnetic symmetry line of the antenna makes it an open-circuit. Hence the impedance increases significantly. This scenario is simulated using the finite element method based (FEM) electromagnetic tool HFSS. Fig. 3(a) illustrates that the resistance becomes almost double as the antenna is halved. Due to the inductive nature of the antenna, the reactive portion of the input impedance exhibits higher inductive values over the band of observation.

Figs. 3(b) and 3(c) illustrates the matching of the both DDD and DHDD antennas with respect to various reference impedances namely, 50, 75, 100, 150 and 200 Ω. It is seen that with 50 Ω reference, DDD shows wide but slightly interrupted -10 dB operating bandwidth; but best performance is achieved (from 2 to 12 GHz) when the reference impedance approximates to 75 Ω. On the other hand, in the case of DHDD, best matching is achieved when the reference impedance is twice the previous value, i.e. 150 Ω. Thus, it is possible to cut an antenna in the magnetic symmetry and adjust the impedance matching by changing its feeding mechanism in the new antenna. It is noted that for all respective impedance references, the resonating frequencies remain unchanged when DDD is cut along the magnetic symmetry line; only the values of reflection coefficients change due to the change in the input impedance.

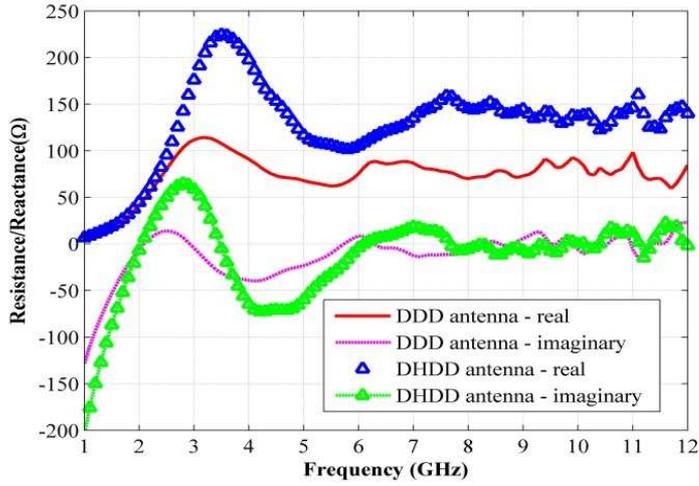

(a)

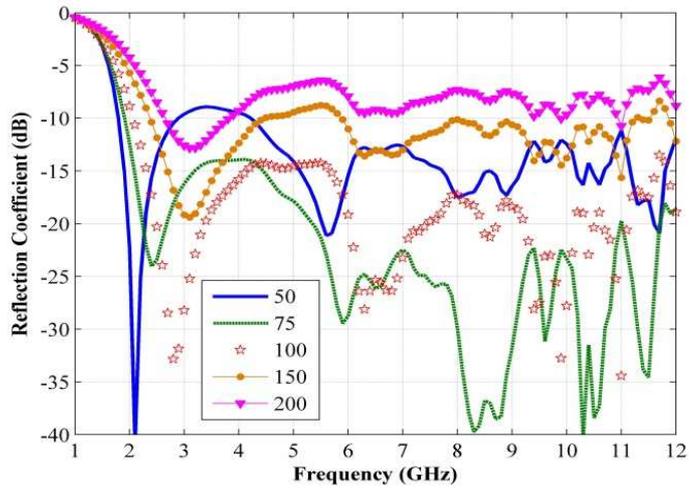

(b)

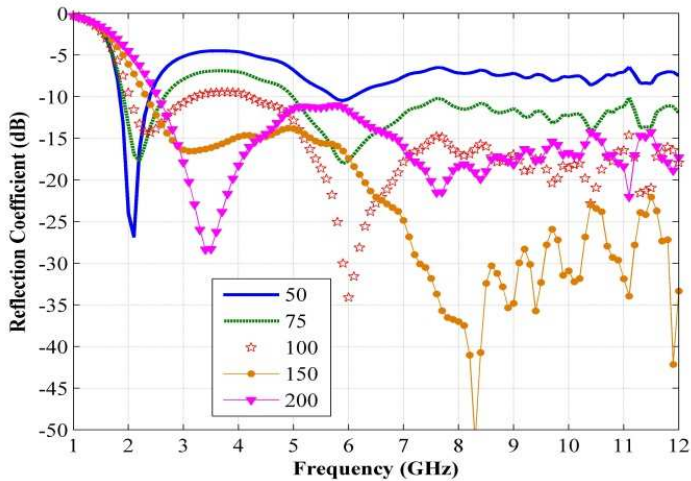

(c)

Fig. 3. (a) Real and imaginary parts of the input impedance of DDD and DHDD antennas; impedance matching of (b) DDD and (c) DHDD antennas with various reference impedances.

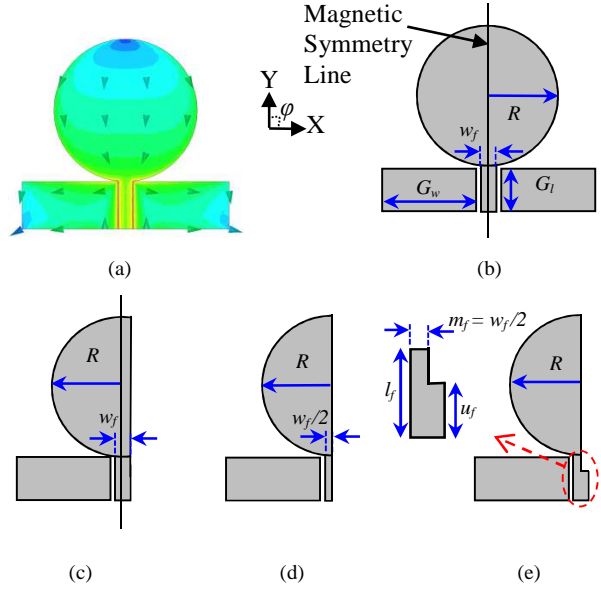

Fig. 4. (a) Surface current distribution and (b) magnetic symmetry condition of conventional monopole antenna; geometric configurations of (c) FUHM, (d) HM and (e) MHM antennas.

### III. Performance Enhancement of Monopole Antennas

It is well known that monopoles are constructed when the electrical symmetry condition of a dipole is met. Fig. 4 shows a conventional monopole (CM) antenna derived from DDD and operating in free-space. The monopole is fed with co-planar ground planes separated by 0.15 *mm* gap from the feeding line which has a width of $w_f = 3$ *mm*. The co-planar grounds are identical to each other with width, $G_w = 20$ *mm* and $G_l = 10$ *mm*. This feeding network is equivalent to 50Ω impedance. Dimensions of the 50Ω microstrip feeding line are calculated using well known microstrip line calculation. For general conceptual relevance, the antenna is analyzed by feeding it with a 50 Ω Sub-Miniature A (SMA) connector. As seen from Fig. 5, the antenna covers a wide bandwidth from 2.8 GHz to more than 11 GHz.

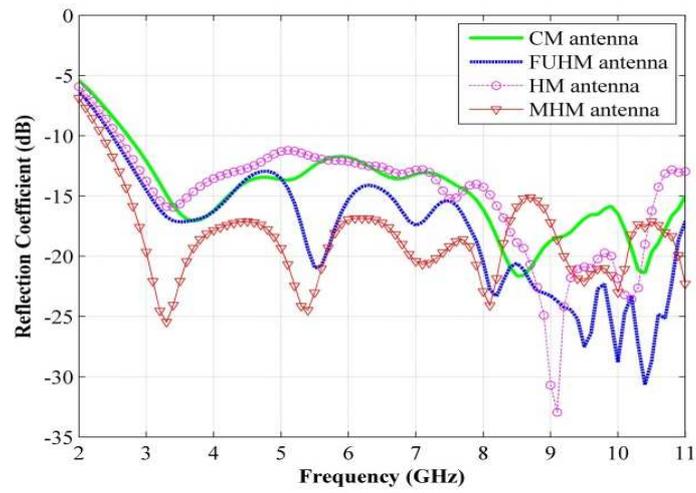

Fig. 5. Impedance matching of monopole antennas shown in Fig. 4 with respect to 50Ω input impedance.

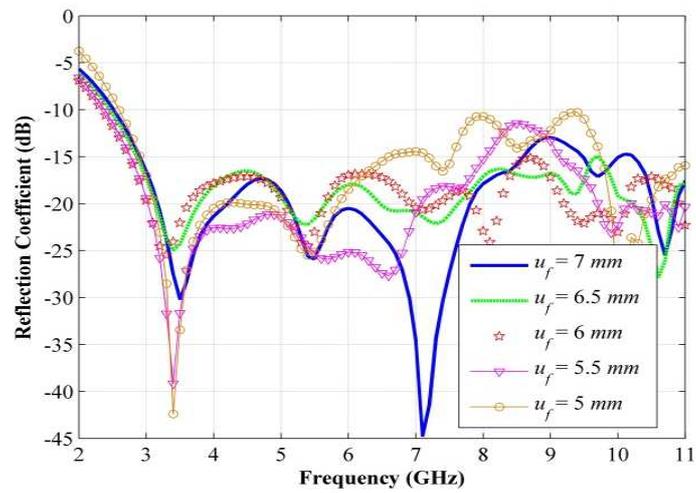

(a)

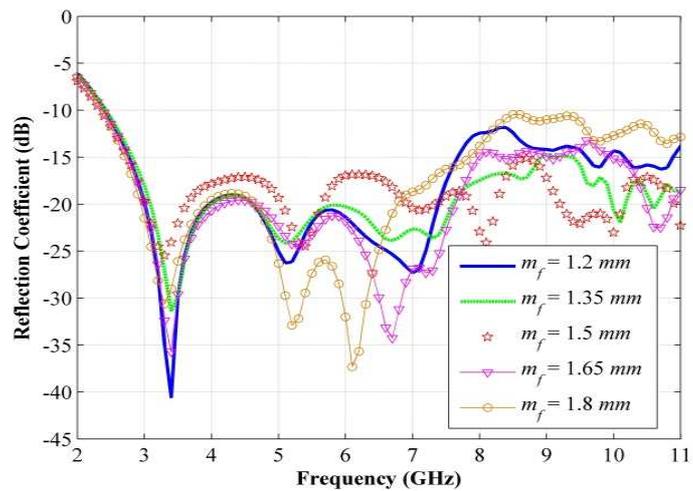

(b)

Fig. 6. Effect on reflection coefficient versus frequency while changing (a) unmodified feeding length ($u_f$) and (b) modified feeding width ($m_f$).

From Fig. 4(a), it is observed that a magnetic symmetry line exists along the Y axis of this monopole. However, unlike the previous DHDD, meeting the symmetry condition also changes the feeding network of the antenna. For this reason, firstly two different quasi-monopoles, namely feed unmodified half monopole (FUHM) and half monopole (HM) antennas are built with slightly different feeding structures (Fig. 4 c, d). The FUHM antenna meets the symmetry condition up to the edge of the feeding line; thus the feeding line is unmodified ($= w_f$) as the CM antenna. However, the HM antenna properly satisfies the symmetry condition; hence the feeding line turns half ($= w_f/2$) of the CM antenna. Using the simulator HFSS, it is seen from Fig. 5 that the antennas achieve better impedance matching performance than the original monopole antenna. These results are in contrary to those of DHDD antenna, which can be attributed to the coupling effect of the ground plane in case of the antennas shown in Fig. 4. Due to the strong capacitive effect of the untouched ground side, the reactive portions of the input impedances of the modified antennas do not exhibit rapid increase of inductance. Moreover, unlike DHDD, due to the symmetrical feeding structure of CM antenna, the modified FUHM and HM antennas attain asymmetrical slot-line feeding network. As a result, the resistive portions of the input impedances of FUHM and HM antennas exhibit slight increments in their values. These combining effects consequently result in a marginal improvement in the impedance matching. Between these two designs, a slightly improved matching is achieved for FUHM when the main feeding line remains unaltered.

Combining these two (FUHM and HM) antennas, another modified half monopole (MHM) is designed (Fig. 4 e). The original feeding line is kept unchanged for $u_f = 6$ $mm$; then the feeding is halved to $m_f = w_f/2$ in order to increase the impedance of the feeding line which connects the halved patch at the end. It is noted from Fig. 5 that MHM antenna attains better impedance matching than the rest of the designs. MHM is basically the optimum derived design from DDD by using HFSS electromagnetic simulator defining $u_f$ and $m_f$ as variables and the other dimensions as constants in order to satisfy both the electrical and magnetic symmetry conditions; while the modified feeding network helps the antenna to reduce the impedance mismatch with the input signal and to achieve a broader radiation band.

It is observed from Fig. 6(a) that changing the unmodified feeding length ($u_f$) has a significant effect on the reflection coefficient performance of the antenna. On the other hand, the lower operating band does not dramatically change when the value of modified feeding width ($w_f$) is varied from the optimized value (Fig. 6(b)). However, the higher operating band is sensitive to the amount of feeding modification. As a general guideline of the modified feeding, it is recommended to maintain a truncated feeding width ratio ($= w_f/m_f$) and truncated feeding length ratio ($= l_f/u_f$) around 0.5. A modal analysis is performed next according to the theory of characteristic modes (TCM) to explain the interest of meeting the electrical and magnetic boundary conditions and the mechanism behind the wider impedance matching of MHM antenna.

### IV. Characteristic Mode Analysis

Modal analysis is done by numerically calculating a weighted set of orthogonal current modes for a given arbitrarily shaped conducting structure. As described in [82], the characteristic modes or characteristics currents are obtained by solving a particular eigenvalue equation that is derived from the Method of Moments (MoM) impedance matrix,

$$X(\vec{J_n}) = \lambda_n R(\vec{J_n})$$

where $\lambda_n$ are the eigenvalues, $\vec{J_n}$ are eigenfunctions or eigencurrents, and R and X are the real and imaginary parts of the MoM impedance operator, $Z = R + jX$.

#### A. Eigenvalue Analysis

Eigenvalues, $\lambda_n$ range from $-\infty$ to $\infty$. The magnitude of a particular eigenvalue represents how efficiently the characteristic mode radiates. Hence, modes with small values of $|\lambda_n|$ radiates efficiently and vice versa. When $|\lambda_n| = 0$, the mode reaches a resonating frequency. Again, if $\lambda_n > 0$, the modes have inductive contribution and stores magnetic energy; while $\lambda_n < 0$, the modes have capacitive contribution and stores electric energy [73]. To analyze the characteristics modes, firstly the eigenvalues are analyzed as they provide information of how the associated modes ($J_n$) radiate and how they are related to the resonance. It is proven that a few modes are required to characterize the mechanism of a small antenna [73, 76].

MoM electromagnetic tool FEKO is proven to implement TCM accurately [83]. Hence, this software is utilized to apply the theory and calculate the characteristic modes. However, the advanced tracking method mentioned in [84] is also adapted to deal with noisy modes and non-continuous behaviour of eigenvalues. Fig. 7(a-c) shows the variation of eigenvalues in different frequencies for the three monopole antenna types, namely the CM, HM and MHM antennas. The modal and resonance characteristics of FUHM are mostly similar to the HM antenna; hence it is not mentioned for the sake of brevity.

The first six eigenvalues are presented here. It is seen that all eigenvalues, except $\lambda_0$, start from the negative side, then crosses the zero level as they reach resonance and continues to increase in the positive side with a gradual slope. The zero eigenvalue, $\lambda_0$ that never reaches the resonance condition ($|\lambda_n| = 0$), holds a positive value in every frequency. This non-resonating mode is associated with the non-resonating current ($\vec{J_0}$) and only contributes to increasing the reactive magnetic power of the antenna. The other modes related to the eigencurrents meet their resonances at some certain frequencies which are listed in table I.

It is seen that the lower resonating frequency primarily depends on $J_1$ mode. As the monopoles are gradually modified from CM to MHM, to optimally match the antenna's input impedance, the resonating frequency of this $J_1$ mode shifts to lower values. This explains the reason behind the enhanced operation of MHM antenna in lower frequencies. In case of CM antenna, eigenvalue analysis shows that mode $J_3$ resonates at 5.50 GHz that is lower than the resonance of mode $J_2$. On the other hand, for HM and MHM antennas, $J_2$ resonates faster than $J_3$ and in a higher frequency compared to mode $J_3$ of CM antenna. It is also noted from Fig. 5 that the second resonance of CM antenna attains lower frequency compared to HM and MHM antennas because of the excitation and dominance of mode $J_3$ of CM antenna on the second resonance.

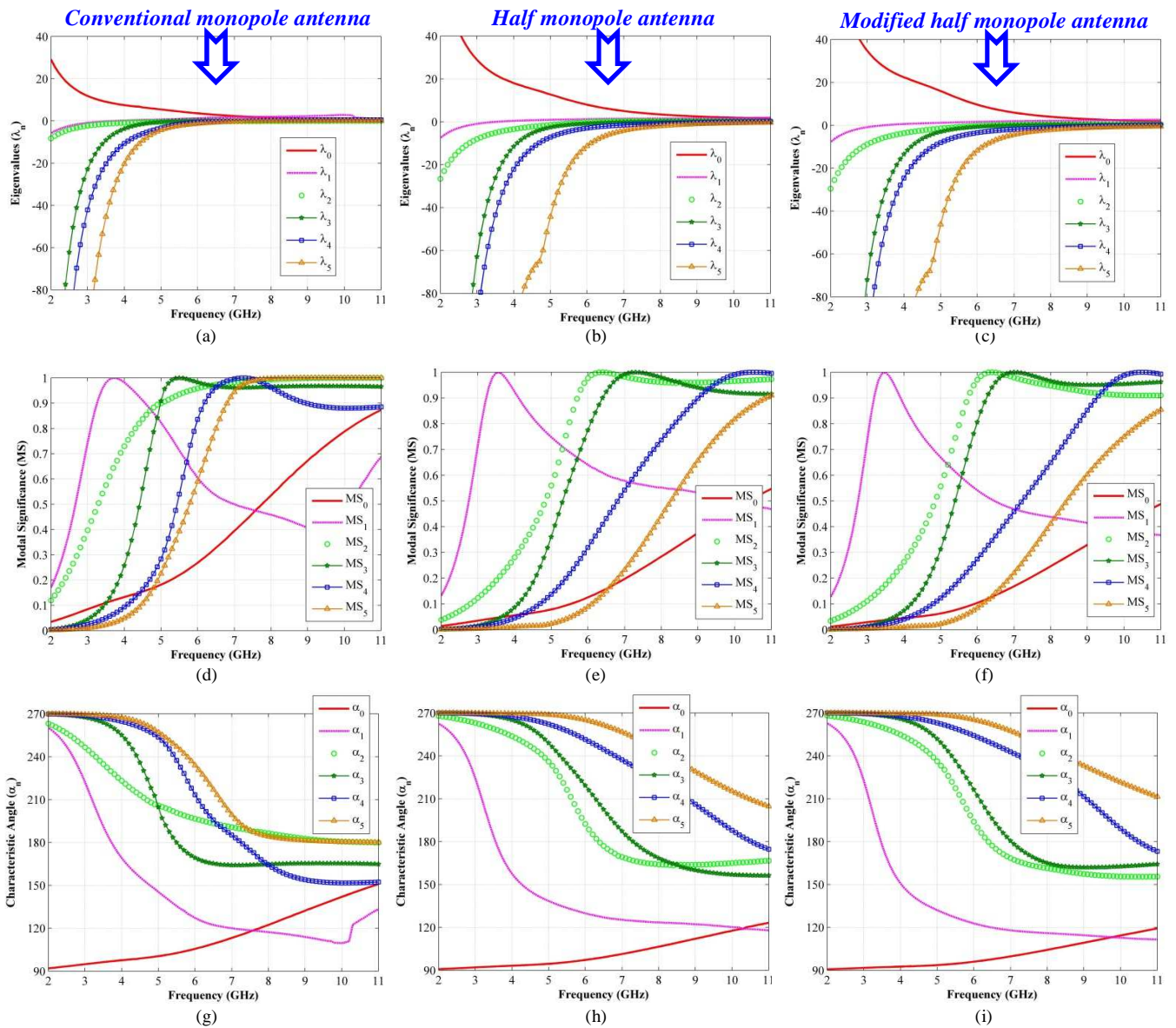

Fig. 7. (a-c) Eigenvalue, (d-f) modal significance, (g-i) characteristic angle curves of conventional monopole, half monopole and modified half monopole antennas respectively.

Table I: Resonant frequency ($f_r$) for the characteristic modes of different monopole antennas

| | Mode $J_1$ | Mode $J_2$ | Mode $J_3$ | Mode $J_4$ | Mode $J_5$ |
|---|---|---|---|---|---|
| **CM** | 3.75 | 10.90 | 5.50 | 7.30 | 10.70 |
| **HM** | 3.55 | 6.45 | 7.3 | 10.55 | - |
| **MHM** | 3.45 | 6.40 | 7.0 | 10.50 | - |

### B. Modal Significance Analysis

In order to highlight the respective roles of characteristic modes, another representation named modal significance ($MS_n$) is preferred over the eigenvalues, which is defined as:

$$MS_n = \left| \frac{1}{1 + j\lambda_n} \right|$$

$MS_n$ represents the inherent normalized amplitude of the characteristic modes [63]. The values of $MS$ state the importance of the respective mode on the scattering characteristics of the antenna. When the value of $MS$ = 1, the mode meets the resonance condition. A mode closer to the highest $MS_n$ curve suggests how effectively that mode is associated to the radiation at a specific frequency.

Figs. 7(d-f) exhibit modal significances of the first six modes for all the three discussed monopole antennas. It is seen that mode zero ($MS_0$) does not resonate over the frequency of operation due to its non-resonating nature. The values suggest that the inductive effect contributed by this mode increases as the modification of the monopole is done from CM to MHM antenna. Similar enhanced inductive effect in the input impedance can also be seen from Fig. 3(a), when the magnetic symmetry condition is considered. The first mode ($MS_1$) is most dominant in lower operating frequencies. It is noted that as the monopole antenna is cut to meet the magnetic symmetry condition, the contributions of other modes become less significant. For a conventional monopole antenna, all the modes have some impact on the operation of the antenna. Hence, the radiation behavior of this antenna is more complex to predict from the current modes. On the other hand, for HM and MHM antennas, the lower band is controlled by exciting mode $J_1$, middle portion by modes $J_2$ and $J_3$, and the last portion by modes $J_4$ and $J_5$. However, since multiple characteristic modes contribute to the high frequency resonances, it is difficult to separate the effects of individual current modes at these higher harmonics.

### C. Characteristic Angle Analysis

The characteristic angle ($\alpha_n$) is another term that is used to explain the operation of an antenna. This angle describes the phase difference between the eigencurrent, $\vec{J}_n$ and its associated characteristic field. It is defined as $\alpha_n = 180° - tan^{-1}(\lambda_n)$. The angle $\alpha_n$ is a measure of the radiation effectiveness of a structure. The modes associated with angle $\alpha_n = 180°$ are at resonance and thus the related structure becomes a good radiator, minimizing stored energy in the reactive near field. However, the modes with $\alpha_n = 90°$ or $270°$ radiates poorly [64, 85]. The nearer the characteristic angle is to $180°$, the better is the radiating manner of the mode.

The characteristic angles of the first six modes of the three investigated antennas are illustrated in Figs. 7(g-i). Although these are other interpretations of Figs. 7(a-c) and 7(d-f), information extraction of resonances is much easier from this plot by merely considering points on $\alpha_n = 180°$ line. It is seen that as the CM antenna is minimized to a halved size by accomplishing the magnetic boundary condition, the lowest-order characteristic mode ($J_1$) resonates in lower frequencies and the slopes become smoother in high frequencies. The middle portion of the operating band of CM antenna is dominated by third ($J_3$) and forth ($J_4$) modes, while the upper band is dominated by $J_2$ and $J_5$. In case of both HM and MHM antennas, the middle and upper parts of the band are influenced by $J_2$ and $J_3$. However, for MHM, these modes resonate in lower frequencies with smoother slopes resulting in better impedance matching over this frequency range.

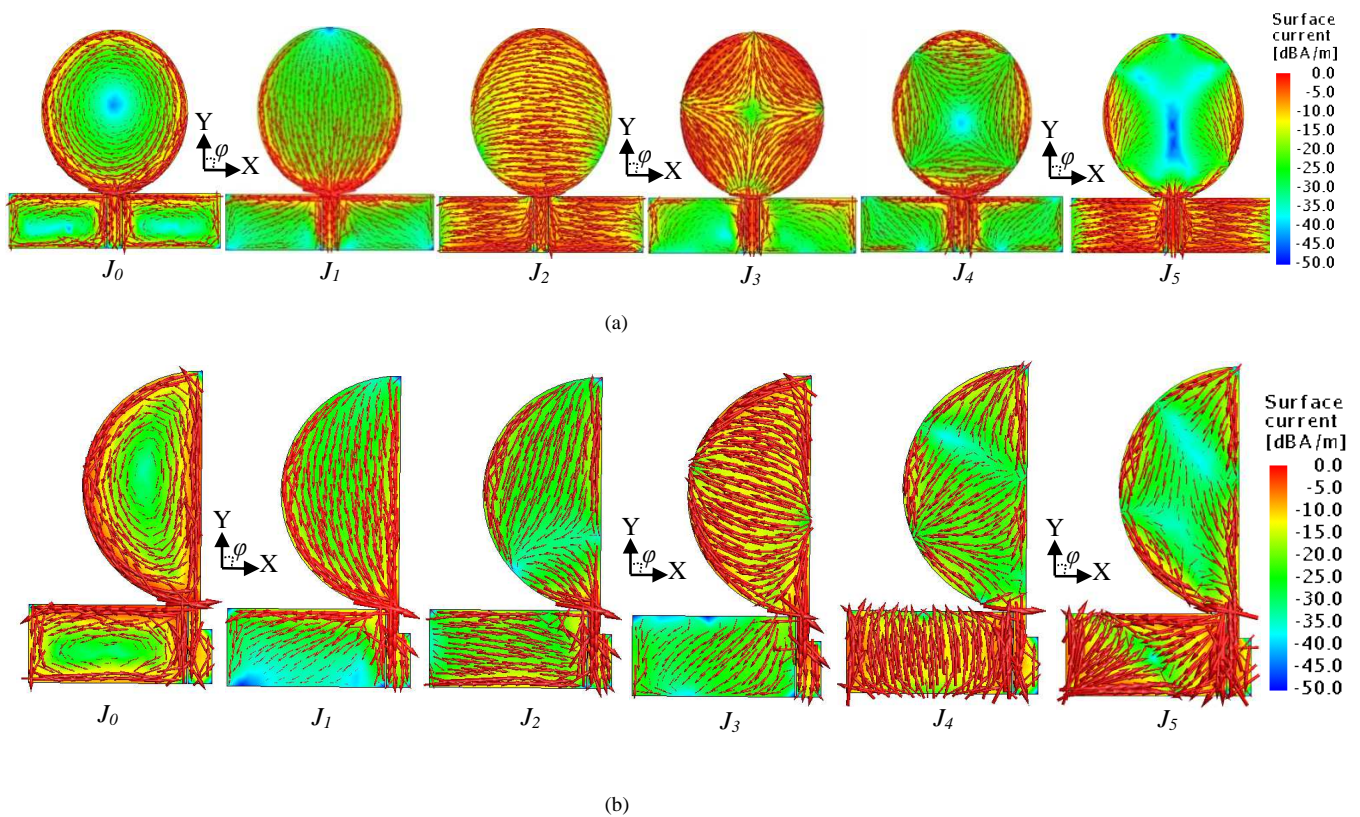

Fig. 8. Normalized magnitude of current distribution at resonance for the first six characteristic modes of (a) CM and (b) MHM antennas.

*D. Eigencurrent Analysis*

The eigencurrents, $\vec{J}_n$ are independent of any specific source or excitation and depend only on the shape and dimensions of the structure and frequency of operation. The total current on the surface of an antenna can be calculated as a sum of $\vec{J}_n$ of the antenna. Similarly, the near and far fields can also be estimated from the independently radiating modal near- and far-fields, respectively. Thus, the modal analysis not only reveals the resonating mechanism of the antenna structure, but also provides an insight into the operation and radiation the antenna.

Fig. 8 exhibits the normalized surface current distributions of the first six characteristic modes of the CM and MHM antennas. All currents contribute in antenna resonances, except mode $J_0$, which is a non-resonating current mode forming a loop like current distribution on each part of the antennas. $J_0$ determines the inductive behaviors of the antennas and helps to store magnetic energy. It should be remembered that these eigencurrent distributions do not rely on the position of the feeding, but on the shape of the antenna. A proper feeding position excites the antenna modes and the exhibit optimum performance of the antenna. Extended simulations show that despite their different modal behavior, the HM antenna exhibits eigencurrent distributions similar to the MHM antenna; hence, they are not presented here.

In CM antenna, most of the currents of mode $J_1$ flow in vertical direction, along Y-axis, and currents of mode $J_2$ flow in horizontal direction, along X-axis. However, in case of mode $J_3$, multiple current nulls are observed with strong current distributions on the circular patch and less currents on the rectangular grounds. It is noted from the discussions of the previous section that most of the operating bandwidth of the antenna is dominated by these three modes. The other two modes, $J_4$ and $J_5$, contribute mostly in the higher frequencies. In case of modes $J_1$, $J_3$ and $J_5$, magnetic symmetry condition is observed along Y-axis. The other modes ($J_0$, $J_2$ and $J_4$) demonstrate electric boundary condition for the CM antenna. However, it is seen for this antenna that the modes ($J_1$, $J_3$ and $J_5$), which exhibit magnetic symmetry, are more significant than the other modes over the operating band (see Fig. 7(a, d, g)). Mode $J_1$ resonates first and is responsible for the lower operating frequencies of CM antennas.

As the CM antenna is halved and the feeding is modified, the proposed MHM antenna meets the magnetic boundary condition. It is seen from Fig. 8 that $J_1$, $J_3$ and $J_5$ modes do not change their characteristic currents after this modification. Most of the currents of these modes still flow vertically along Y-axis. Mode $J_0$ creates a loop shape in the half circular patch. Currents on the remaining rectangular ground do not alter their directions. A significant change of current flow is observed in mode $J_2$. Most of the currents flow vertically, starting from the current null. As described later, vertical currents plays an important role on the radiation patterns of the antenna. Since mode $J_2$ resonates in a quicker manner than that of CM antenna, these currents help MHM antenna to improve its radiation characteristics over 5 to 7 GHz. In case of MHM antenna, the resonances of modes $J_4$ and $J_5$ shift to higher frequencies. Thus, these modes have lesser effect on MHM antenna performance than CM antenna.

Since the modal currents are independent to antenna excitation, suitable feeding position of antennas is critically important to excite specific characteristic modes. It is also noted from Fig. 8 that the CPW strip line consistently carries strong current vectors in both the structures of CM and MHM antennas. The areas having large current magnitude in regular manner, suggest that feeding the antennas from the strip line end will excite the eigencurrents and thus, the characteristic modes properly [86].

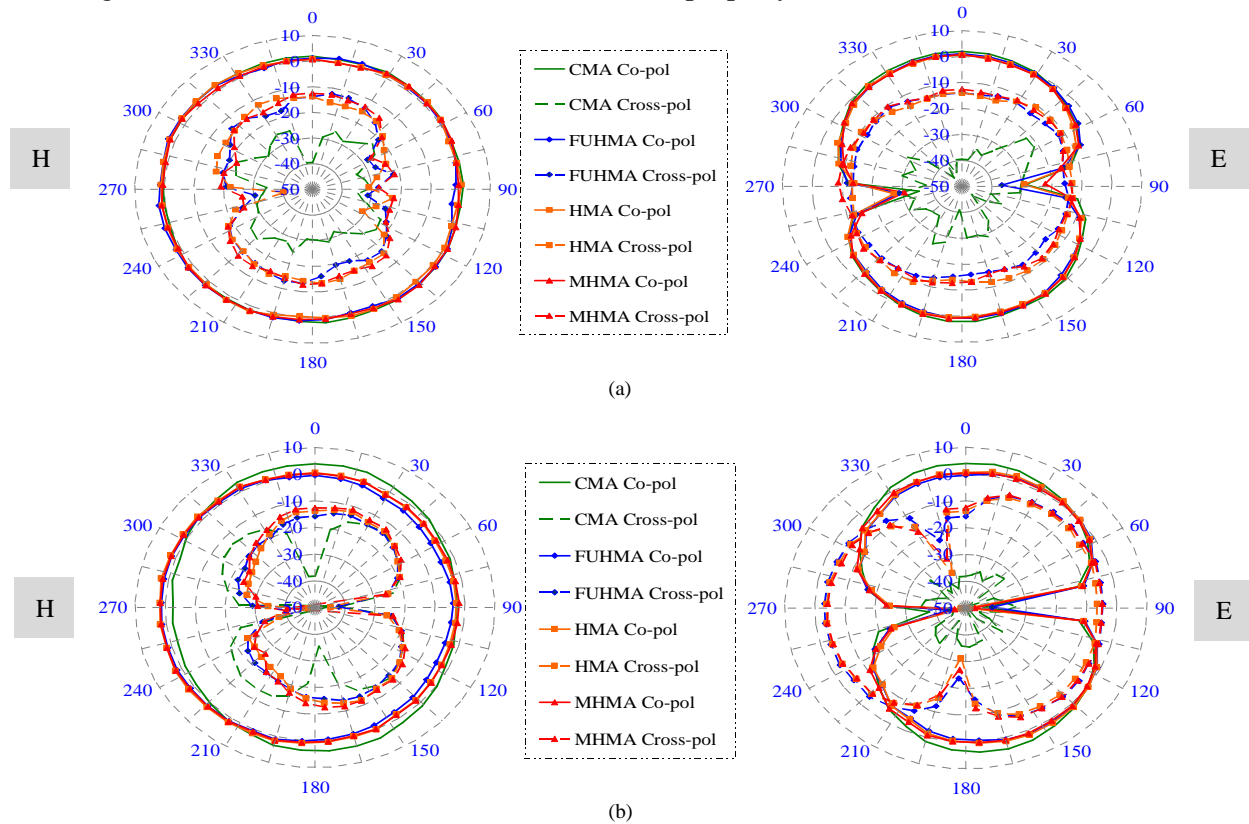

Fig. 9. H-(XZ) and E-(YZ) plane pattern of various monopole antennas extracted from HFSS at (a) 4 and (b) 8 GHz frequency.

### E. Correlation between eigencurrents and radiation patterns

The radiation patterns of all four quasi-monopole antennas at 4 and 8 GHz are illustrated in Fig. 9. Good omnidirectional radiation patterns in H-(XZ) plane are achieved for all the antennas. In spite of the asymmetrical structures of the halved monopoles, they exhibit symmetrical radiation patterns in lower frequencies. With the increase of operating frequency, as the relative sizes of the asymmetric ground planes increase, the antennas tend to radiate slightly more along the X-direction. The main direction of radiation (Z-direction) still remains relatively high. Similar phenomena are observed in case of E-(YZ) plane radiation. However, for both planes, a significant increase in cross-polarization level is observed. This feature can be explained from the eigencurrents shown in Fig. 8, as the field distribution fully depends on the variation of current strength and direction.

Mode $J_1$ is mostly responsible for the lower operating frequencies of the monopole antennas. A strong current flows on the circular patch along Y-axis, while the rectangular ground carries the horizontal current components. It is worth to mention that since the antennas are fed from a CPW feeding point, the vertical (Y-axis) currents are responsible for the co-polar radiation. The horizontal (X-axis) currents contribute in building up the cross-polarization of the radiation patterns. For CM antenna, the X-axis currents are out of phase along Y-axis, creating a magnetic symmetry line. Hence, the fields generated by these currents cancel each other limiting the cross-polarization level to relatively low. On the other hand, for the halved monopoles the counter radiation of the horizontal currents is absent due to the size reduction. Hence, these horizontal currents create high cross-polarization level for the halved monopoles.

Mode $J_2$ is more significant in the halved antennas than in the conventional monopole antenna. The altered vertical currents of mode $J_2$ reduces the effect of increased horizontal currents in the middle portion of the band. The same mode also affects the antenna's radiation in higher frequencies along with mode $J_3$.

Although the impedance matching of the proposed MHM antenna improves quite well than the other halved antennas (FUHMA and HMA), they show similar radiation patterns and cross-polarization level. It is recommended that decreasing the width of the rectangular ground decreases the horizontal currents and, thus, the cross-polarization level. However, slight deficiency in the operating frequency band is observed.

The peak gain plots of the four antennas are illustrated in Fig. 10. The gain increases gradually with the increase in operating frequencies. A slight variation in the gain is observed from one antenna to another. Although the radiation slightly decreases along Z-direction for halved antennas, the peak gain level is still comparable to the CM antenna.

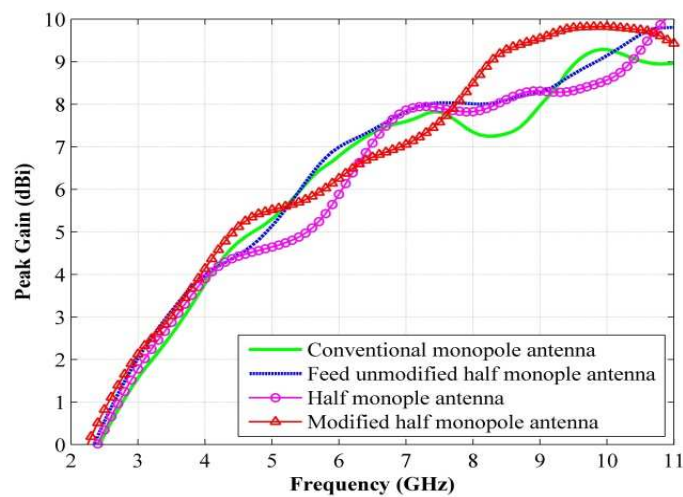

Fig. 10. Maximum gain patterns of different monopole antennas.

## V. Design Examples

### A. CPW Fed Quasi-monopole Antenna

The proposed modified feeding technique for half monopole antennas is applied for the miniaturization of compact CPW fed quasi-monopole antenna described in [87]. That antenna provides UWB performance by loading inverted L-strip over typical monopole radiator on one side of 1.6 mm thick FR4 substrate (permittivity of 4.4 and loss tangent of 0.024). The original design and our modified variations are illustrated in Fig. 11. All the dimensions given in this and subsequent antenna structures are in (mm). The FUHM and HM antennas are shaped by chopping half of the antenna while keeping the full and half of the feeding line respectively. The FUHM, HM and MHM antennas are designed following the findings of the previous discussions on the analysis of characteristics modes based on the antennas of Fig. 4. In the MHM antenna formation, the original feed remains up to 5.3 *mm* length and then the feed is cut into half of the width and extended and connected to the radiating patch 8.8 *mm* from the SMA connection point.

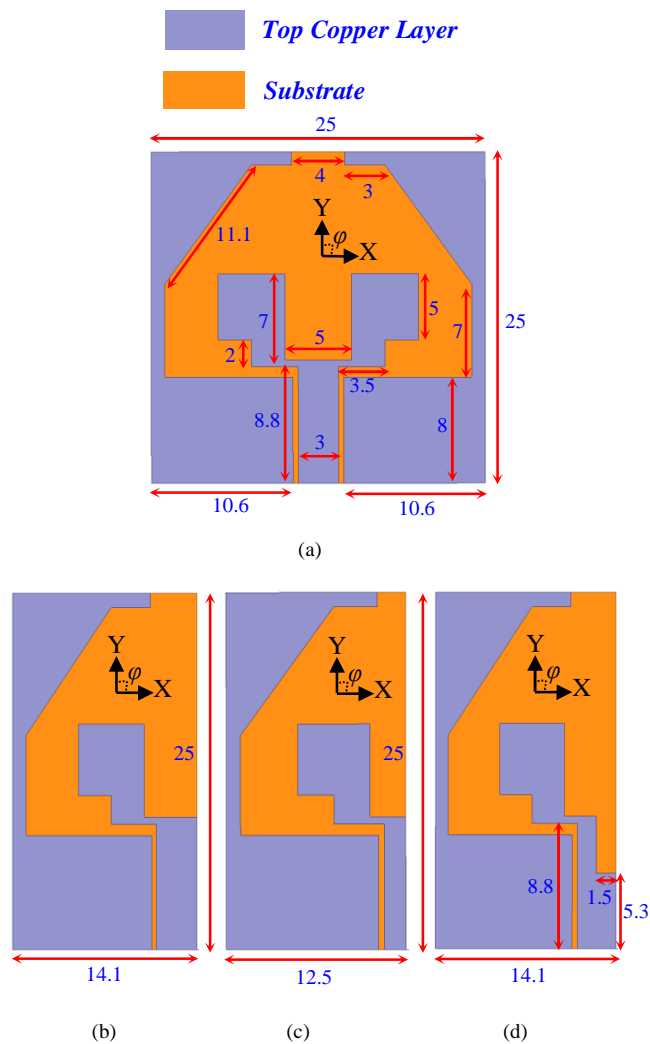

Fig. 11. Schematic diagrams of (a) CM, (b) FUHM, (c) HM and (d) MHM version of the antenna in [87] .

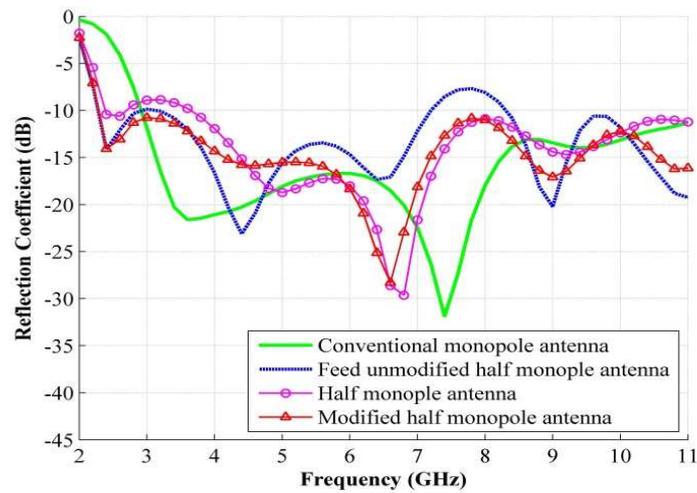

Fig. 12. Impedance matching of antennas shown in Fig. 11 with a 50Ω feed.

The results of the antenna miniaturization extracted from HFSS, are exhibited in Fig. 12. The miniaturized antennas (FUHM and HM) provide wider bandwidths than the conventional design. However, they lack in maintaining proper impedance matching level and impose instability. The MHM antenna, on the other hand, clearly overcomes these problems; it supports radiation from 2.2 GHz to over 11 GHz with respect to –10 dB reflection coefficient reference line. A 14% increase in impedance bandwidth is achieved in MHM antenna as a result of the new miniaturization technique with a 44% size reduction from the original quasi-monopole antenna.

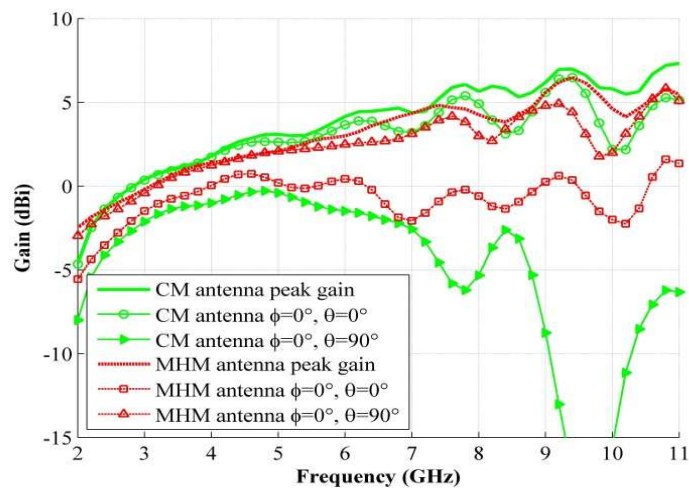

Fig. 13. Gain patterns of CM and MHM antennas of Fig. 11 in various directions.

Comparing the gains of CM and MHM antennas provides an insight into the change in the radiation characteristics due to the miniaturization. As shown in Fig. 13, the peak co-polar gain attained from MHM antenna slightly decreases compared to that of CM antenna due to the reduction in the radiating area. The negative gains at lower frequencies are due to the material loss of FR4 substrate which can be overcome by using low loss substrates. The co-polar gain curve of the MHM antenna along Z-axis direction ($\phi=0°$, $\Theta=0°$) tends to roam around 0 dBi, while for CM antenna, the gain matches well with peak gain as the CM antenna radiates better than the MHM antenna along this direction due to the symmetry of the CM antenna. However, it is noted that the co-polar gain plot of the MHM antenna along Y-axis direction ($\phi=0°$, $\Theta=90°$) corresponds to the maximum gain curve well. This concludes that the MHM antenna radiates mainly along $\phi=0°$, $\Theta=90°$ direction. Comparing to the CM antenna, the gain values are much higher in this direction. If omni-directionality is of concern, it can be observed by looking at the gain curves that MHM antenna radiates better than the CM antenna in this manner.

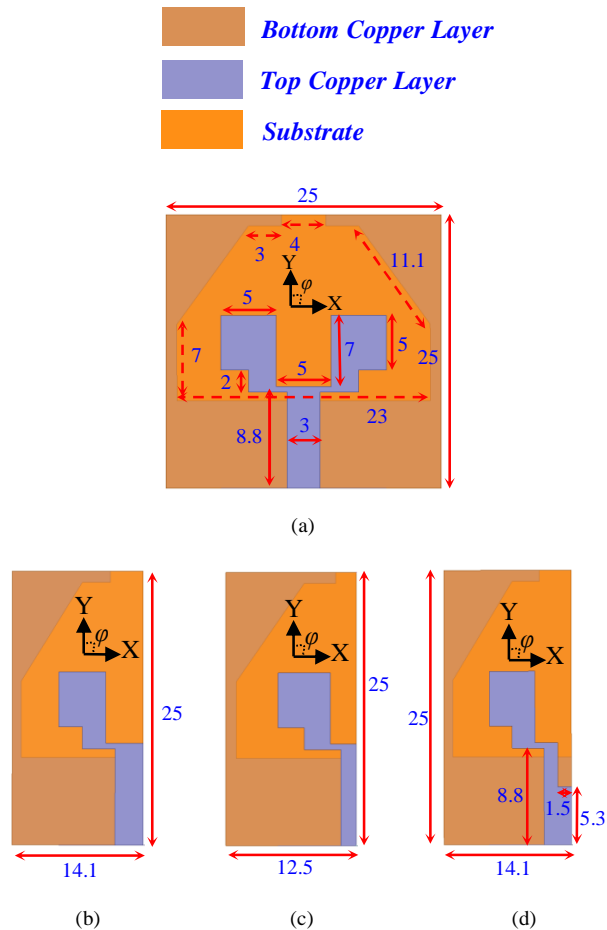

Fig. 14. Geometric configurations microstrip feed modifications of (a) CM, (b) FUHM, (c) HM and (d) MHM antennas shown in Fig. 11.

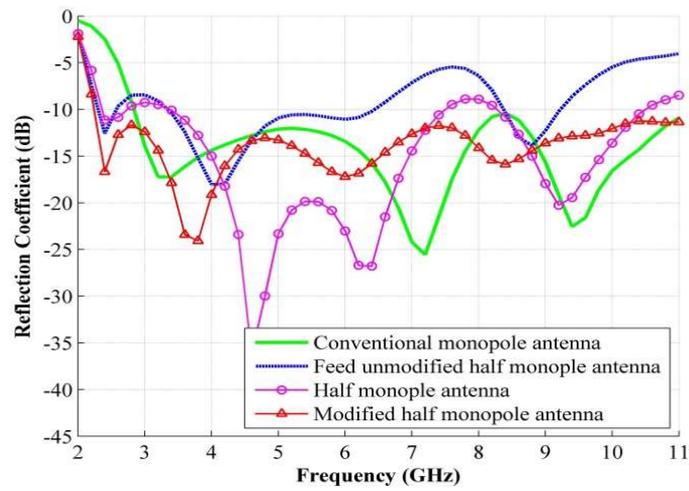

Fig. 15. Reflection coefficients of the antennas of Fig. 14.

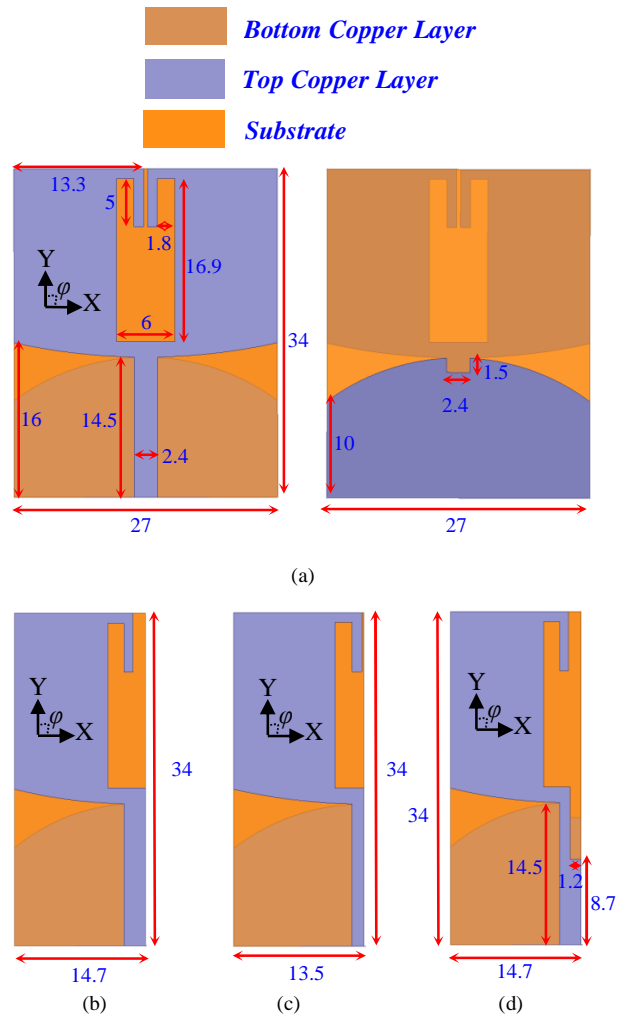

Fig. 16. Top and bottom views of microstrip fed (a) CM, (b) FUHM, (c) HM and (d) MHM antennas, modified from [38].

## B. Microstrip Line Fed Quasi-monopole Antenna

The miniaturization technique is also applied to the microstrip fed antennas. In the first example, the aforementioned CPW antenna is converted to a microstrip fed UWB antenna and exhibited in Fig. 14. After simulating the structures using HFSS, similar characteristics were seen as antecedently discussed (Fig. 15). For the sake of brevity, the explanations are not repeated.

For the second example, a top capacitively-loaded loop UWB antenna [38] is chosen for the verification. This microstrip fed antenna is fabricated on 0.787 *mm* thick Rogers Duroid 5880 material with permittivity of 2.2 and loss tangent of 0.0009. The geometry of the antenna and the modified variations are exhibited in Fig. 16. The final MHM version of this antenna consists of half of the original structure along with the modified feeding which comprises of 8.7 *mm* long original feed line which is then truncated to half to feed the radiating patch. A truncation ratio of 0.6 is maintained in this case. It is seen from Fig. 17 that the HM antenna provides wider bandwidth than CM antenna with good impedance matching, but presents narrower operating band than MHM version. The MHM antenna covers an operating band from 2.4 to more than 11 GHz, which is around 10% fractional bandwidth improvement with 46% size reduction compared with the original CM edition.

Table II summarizes the results of the proposed miniaturization technique for both CPW and microstrip fed symmetrical antennas. Distinct advantages are seen vividly upon the application of symmetry conditions and generalized modified feeding method on both types of antennas.

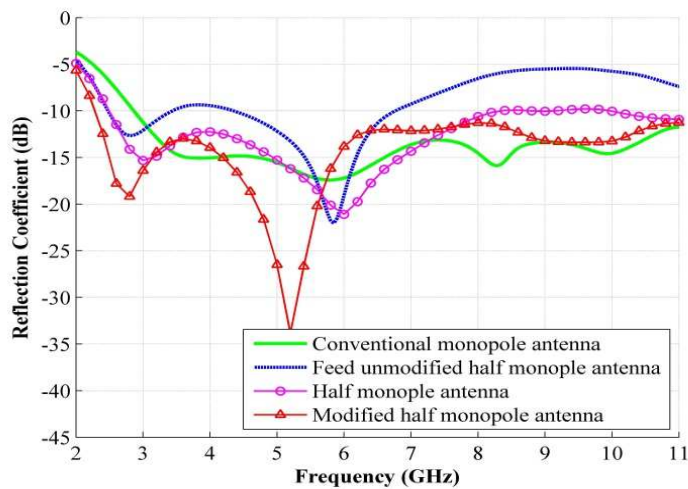

Fig. 17.   Impedance matching characteristics of different antennas of Fig. 16.

Table II: Summary of applying the proposed miniaturization techniques on various antennas

| Antennas | [87] | Microstrip Fed [87] | [38] |
|---|---|---|---|
| Truncated width ratio ($w_f / m_f$) | 1.5/3 = 0.5 | 1.5/3 = 0.5 | 1.2/2.4 = 0.5 |
| Truncated length ratio ($l_f / u_f$) | 5.3/8.8 = 0.61 | 5.3/8.8 = 0.61 | 8.7/14.5 = 0.6 |
| Lowest operating frequency of CM antenna (GHz) | 2.85 | 2.8 | 2.9 |
| Lowest operating frequency of MHM antenna (GHz) | 2.2 | 2.25 | 2.4 |
| Fractional Bandwidth of CM antenna (%) | 117 | 118 | 116 |
| Fractional Bandwidth of MHM antenna (%) | 133 | 132 | 128 |
| Size reduction from CM to MHM antenna (%) | 44 | 44 | 46 |

## VI. Conclusion

With the invention of printed circuit board (PCB) techniques, extensive research has been done to achieve wider bandwidth, low cost, compact and planar antenna designs, in past few decades. As a result, a huge number of ultra-wideband (UWB) antennas are reported. Most of the existing compact planar designs are monopole antennas with symmetrical structures which are derived by meeting the electrical symmetry of their dipole version. In this paper, a new half cut monopole antenna technique has been proposed by meeting the magnetic symmetry condition and feeding network modification. The modified feeding technique can be used as a general guideline to utilize the existing symmetrical designs for compact structures and wider bandwidths. In order to look deep into the physics behind the enhanced performance, modal analysis is adopted. The main reason attributed to the effects of the modified structures is identified. It has been shown that the modes responsible for modified half monopole (MHM) antenna's operation resonate in lower frequencies with smoother slopes than the other feed unmodified half monopole (FUHM) and half monopole (HM) antennas. The excitation independent characteristic currents (eigencurrents) are also observed by the theory of characteristic modes and correlated to the radiation patterns of MHM antennas. It is seen that fields created by the horizontal currents, which loses their opposing counterpart from other half of the antenna, are responsible for the increased cross-polarization level. However, it is seen that the MHM antenna radiates in a similar manner as FUHM and HM antennas. Finally, some examples of the proposed feeding techniques have been illustrated for both coplanar waveguide and microwave feeding. Using the modified half monopole (MHM) structures, a fractional bandwidth improvement of 10-15% is accomplished with a size reduction of around 45%. It is also observed that the MHM radiates in a more omni-directional manner than their conventional monopoles counterparts. The new feeding technique will be of help to design compact UWB antennas for chip integrated transceivers.

<div align="center">REFERENCES</div>